# Depletion-to-Enhancement Mode Transition and Strongly Suppressed Hysteresis in Surface Engineered Multilayer $MoS_2$ FETs

Samiksha Bhatia[1], Ramesh Singh Bisht [1], and Pramod Kumar[1,*]
[1]*Department of Physics, Indian Institute of Technology Bombay, Mumbai 400076, India*

[*]Corresponding author: pramod_k@iitb.ac.in

**Abstract:** Two-dimensional (2D) semiconductors such as molybdenum disulfide ($MoS_2$) have recently attracted extensive research attention due to their promising compatibility with silicon-based electronics. However, several key challenges still limit their practical integration. Two of the critical issues are: (1) the intrinsic depletion-mode (normally-on) operation of $MoS_2$ field-effect transistors (FETs), and (2) the large hysteresis commonly observed in the transfer characteristics of $MoS_2$ FETs due to the inherent sulfur defects. Addressing them is essential for CMOS-compatible 2D-transistor technologies. In this work, we report for the first time that surface modification of the exfoliated multilayer $MoS_2$ FETs with PBTTT-C14 (poly(2,5-bis(3-tetradecylthiophen-2-yl)thieno[3,2-b]thiophene)), a p-type conjugated organic polymer, converts the device from depletion-mode to enhancement-mode operation while simultaneously and strongly suppressing hysteresis. Specifically, the threshold voltage ($V_{th}$) shifts from −9.6 V to +5.9 V (total shift ~15.5 V), and the hysteresis window decreases from 8.8 V to 1.3 V (~85% reduction). This originates from interfacial charge transfer at the $MoS_2$/PBTTT-C14 interface, enabled by favourable band alignment. Additionally, donor-like sulfur vacancy states in $MoS_2$ are involved in interfacial charge transfer to the p-type PBTTT-C14 layer, resulting in effective deactivation of these donor states associated with the generation of free charge carriers and hysteresis. Importantly, this highlights the role of electronic interaction at the interface. To further validate this charge-transfer-driven mechanism, P3HT (poly(3-hexylthiophene-2,5-diyl)) with similar energy levels to PBTTT-C14 was employed, and it also showed similar enhancement-mode behaviour and hysteresis suppression. In contrast, treatment with a n-type polymer (N2200) (poly{[N,N′-bis(2-octyldodecyl)-naphthalene-1,4,5,8-bis(dicarboximide)-2,6-diyl]-alt-5,5′-(2,2′-bithiophene)}) with less favourable band alignment which limits interfacial charge transfer, resulted in a smaller $V_{th}$ shift from −13.8 V to −8.3 V (total shift of ~5.5 V) without inducing enhancement-mode behaviour and a hysteresis reduction from 11.2 V to 5.6 V (~50%). Furthermore, treatment with the high band gap, insulating polymer polydimethylsiloxane (PDMS) leads to no observable change in $V_{th}$ and hysteresis, supporting the charge-transfer-dominated origin of the observed effects. Overall, this work establishes a clear correlation between polymer energy-level alignment, interfacial charge transfer, and device performance. The results demonstrate that p-type organic polymers with suitable energy levels can simultaneously enable enhancement-mode operation and strong hysteresis suppression in $MoS_2$ FETs, providing a scalable and CMOS-compatible strategy for high-performance 2D electronics.

## Introduction

As predicted by Moore's law, the number of transistors on an integrated chip doubles approximately every two years, driving continuous efforts to miniaturize device dimensions[1]. Silicon technology has already reached nodes below 5 nm, but further downscaling faces fundamental limits due to short-channel effects, leakage currents, and reliability issues[2]. Consequently, the search for alternative materials capable of sustaining Moore's law beyond the silicon era has become an urgent priority[3]. Following the discovery of graphene in 2004[4], two-dimensional (2D) materials have emerged as promising candidates for next-generation electronics. Although graphene exhibits exceptional carrier mobility (up to 5000 $cm^2\ V^{-1}\ s^{-1}$), its zero bandgap leads to extremely low on/off ratios (~$10^2$), making it unsuitable for logic circuit applications[5,6]. Transition metal dichalcogenides (TMDs), on the other hand, possess tunable bandgaps, high on/off ratios, excellent mobility, mechanical flexibility, and chemical stability, as well as strong light–matter interactions and compatibility with flexible electronics[7–9]. Among them, molybdenum disulfide ($MoS_2$) has been most extensively studied for over a decade owing to its natural abundance, environmental stability and favourable electronic properties[10–12]. Significant efforts are underway to achieve CMOS-compatible $MoS_2$-based electronics. However, several challenges remain to be resolved, such as achieving large-area uniform growth[13,14], developing suitable contacts and dielectrics[15,16], precise threshold voltage ($V_{th}$) control[17,] achieving enhancement-mode (E-mode) operation[18], hysteresis suppression[19], development of effective passivation layers with minimized interface traps[20], and implementing p-type doping strategies[21]. Among these, this work particularly addresses two of the most crucial challenges, achieving E-mode operation and hysteresis suppression, which are key prerequisites for low-power and reliable logic devices. $MoS_2$ FETs typically operate in the depletion-mode (D-mode) due to intrinsic n-type doping[22], leading to a negative $V_{th}$. Such normally-on behaviour is undesirable for logic circuits, as it requires additional gate bias to turn off the device, thereby increasing static power consumption. Further, in $MoS_2$ based FETs, a large hysteresis is often observed[23]. It is the difference between forward and backward gate-voltage sweeps, arises from charge trapping/detrapping processes associated with surface adsorbates, interfaces and defects within the channel[24–26], this leads to instability and reliability concern in electrical performance. In conventional silicon-based transistor technology, $V_{th}$ modulation is easily achieved through ion implantation. However, in 2D materials like $MoS_2$, such doping is not possible due to their 2D atomically thin nature[21]. Recent studies have thus explored alternative routes to enable E-mode operation utilizing various technique. For instance, Wang *et al.* demonstrated E-mode $MoS_2$ FETs through gate metal work function engineering[27]; Yoo *et al.* achieved E-mode operation by employing a fluorine-containing gate dielectric (CYTOP)[28]; Leong *et al.* demonstrated bidirectional control of the $V_{th}$ through sulfur and hydrogen surface treatments[29]; Kang *et al.* demonstrated $V_{th}$ tuning by using $NH_3$ plasma treatment[30]; Roh *et al.* realized E-mode operation using self-assembled monolayer (SAM) modification with octadecyltrichlorosilane (OTS)[31]; Rai *et al.* demonstrated E-mode behaviour through argon plasma treatment followed by $O_2$ bath[18]; and Li *et al.* reported $V_{th}$ modulation by employing different organic solvents for surface modification[32]. Consequently, surface engineering has emerged as the most promising approach for tuning the $V_{th}$ in 2D materials, as it precisely modulates the channel potential without introducing scattering centres or compromising lattice integrity. However, this field remains actively developing, and several critical challenges persist. In most of the studies, the total $V_{th}$ shift is limited to a few volts, while most of the pristine $MoS_2$ FETs in literature typically exhibit highly negative $V_{th}$, making a large positive $V_{th}$ shifts essential. Additionally, in the above studies, a critical reliability metric, i.e., hysteresis behaviour after treatment, remains unaddressed, and moreover, the temporal stability of $V_{th}$ over extended periods has not been investigated. Parallelly, significant efforts have also been directed toward minimizing hysteresis in $MoS_2$ FETs. For example, Haa *et al.* demonstrated a ~53% reduction in hysteresis using an APTES passivation layer, however no significant change in $V_{th}$ is observed, thereby leaving the device in depletion-mode[33]; Liu *et al.* demonstrated that $O_2$ plasma followed by an $Al_2O_3$ passivation layer reduces hysteresis by nearly 94%, with no significant change in $V_{th}$[34]; Kang *et al.* reported a ~27% reduction in hysteresis accompanied by a negative shift in $V_{th}$ with $CsPbBr_3$ nanoclusters[35]; Jana *et al.*

achieved 96% hysteresis reduction using a hydrophobic organosilane HMDS treatment, although $V_{th}$ shifted slightly in the right direction, it remained strongly negative, leaving the device in depletion-mode[36]. Collectively, these studies highlight the following critical gap; (1) achieving a large, positive and stable $V_{th}$ shift (sufficient for D→E conversion), and (2) simultaneously addressing hysteresis, or ideally, minimizing it to ensure reliable device performance. In this context, organic polymers offer a compelling alternative. Owing to the absence of dangling bonds and native surface oxides, they interact with 2D materials primarily through van der Waals forces, forming a gentle, non-destructive interface that does not introduce additional trap states and preserves the structural integrity of $MoS_2$[37–40]. Their tunable energy levels, together with the ability to incorporate diverse functional groups, enable precise interfacial charge-transfer modulation and controlled tuning of $V_{th}$[41–43]. Moreover, their mechanical flexibility makes them highly suitable for surface engineering and integration into next-generation flexible electronics[44]. Here, we demonstrate for the first time that employing PBTTT-C14 (poly(2,5-bis(3-tetradecylthiophen-2-yl)thieno[3,2-b]thiophene)), a p-type conjugated hydrophobic polymer, enables E-mode operation in multilayer $MoS_2$ FETs, shifting the threshold voltage ($V_{th}$) from a large negative value (–9.6 V) toward positive value (+5.9 V) (total shift ~15.5 V), effectively converting the device from D-mode to E-mode operation, while simultaneously achieving strong hysteresis suppression from 8.8 to 1.3 V (~ 85 %). The modulation arises from interfacial charge-transfer due to favourable energy band alignment; PBTTT-C14 extract electrons from $MoS_2$, lowering its Fermi level, thus inducing a positive shift in the $V_{th}$. A second p-type polymer, P3HT (poly(3-hexylthiophene-2,5-diyl)), with a comparable energy level to PBTTT-C14, produces similar results. In contrast, coating with the n-type polymer N2200 (Poly{[N,N′-bis(2-octyldodecyl)-naphthalene-1,4,5,8-bis(dicarboximide)-2,6-diyl]-alt-5,5′-(2,2′-bithiophene)}) yields a smaller $V_{th}$ shift (~ 5.5 V) without E-mode conversion and moderate hysteresis reduction (~50 %) due to minimal charge-transfer due to unfavourable band level alignment. Furthermore, the ambient passivation layer of insulating polymer PDMS (Polymethyl siloxane), shows no observable shift in $V_{th}$ and hysteresis, confirming how passivation from ambient gases results in no improvement. These control experiments demonstrate that significant shifts in $V_{th}$ and hysteresis suppression require substantial interfacial electronic interaction and charge transfer; a merely electronically inert overlayer and passivation cannot induce significant modulation in $V_{th}$ and hysteresis. Thus, polymers such as PBTTT-C14 and P3HT, with favourable energy-level alignment to $MoS_2$, serve as effective interfacial layers for the simultaneous control of both the parameters.

Additionally, the temporal stability of $V_{th}$, monitored over more than 10 days, and the consistent trends observed across multiple devices (~10 samples) confirm the reproducibility and reliability of this approach. This study thus establishes a robust strategy that enables large and stable threshold voltage ($V_{th}$) control and simultaneously and strongly supressing hysteresis through a single-step, low-damage process. Such effective control over interfacial properties paves the way for tunable, low-power, and reliable 2D semiconductor transistors, with promising implications for future CMOS logic, flexible electronics, and neuromorphic systems.

## Experimental

Multilayer $MoS_2$ flakes were mechanically exfoliated from a bulk $MoS_2$ crystal (purchased from Ossila, Solpro Business Park, UK) using adhesive tape and a polydimethylsiloxane (PDMS) stamp, and transferred onto a Si substrate with 200 nm thermally grown $SiO_2$. Back-gated FETs were fabricated using optical lithography and sputtering techniques, where highly doped p-type silicon served as the back gate, and Ni/Au (30/100 nm) contacts were deposited as source and drain electrodes. Organic polymer powders (PBTTT-C14, P3HT, and N2200; purchased from Sigma-Aldrich, Inc.) were dissolved by adding 5 mg of each to 1 mL of chlorobenzene, followed by magnetic stirring at 400 rpm for 1.5 h to obtain homogeneous solutions. The polymer solutions were spin-coated onto the devices in a two-step process (250 rpm for 10 s, then 1000 rpm for 30 s), followed by thermal annealing at 150 °C for 1 h to promote adhesion and remove residual solvent. Prior to electrical measurements, the polymer layer was gently removed from the probe-contact regions using chlorobenzene. PDMS films were prepared by dissolving PDMS gel in a 10:1 ratio with the curing agent, followed by drop-casting onto the device and thermal annealing at 150 °C for 1 h to ensure film stabilization. The thicknesses of the $MoS_2$ flakes and resulting polymer layers were measured using atomic force microscopy (AFM; Oxford Instruments Asylum Research, MFP-3D BIO, USA). Optical images were obtained with an Olympus MX61-F microscope. Electrical characterizations were performed using a Polaris B1500 Agilent Semiconductor Characterization System under ambient conditions at room temperature.

## Result and discussion:

A back-gated multilayer $MoS_2$ FET was first fabricated and electrically characterized to establish the reference behaviour of the pristine device. After this, a thin film of PBTTT-C14, a p-type conjugated polymer (poly[2,5-bis(3-tetradecylthiophen-2-yl)thieno[3,2-b]thiophene]) was spin-coated onto the $MoS_2$ channel, following the procedure described in the Experimental Section. A schematic illustration of the PBTTT-C14 coating process is presented in Figure 1(a to c). The chemical structure of PBTTT-C14 and the optical image of the fabricated device are shown in Figure 1(d) and Figure 1(e), respectively.

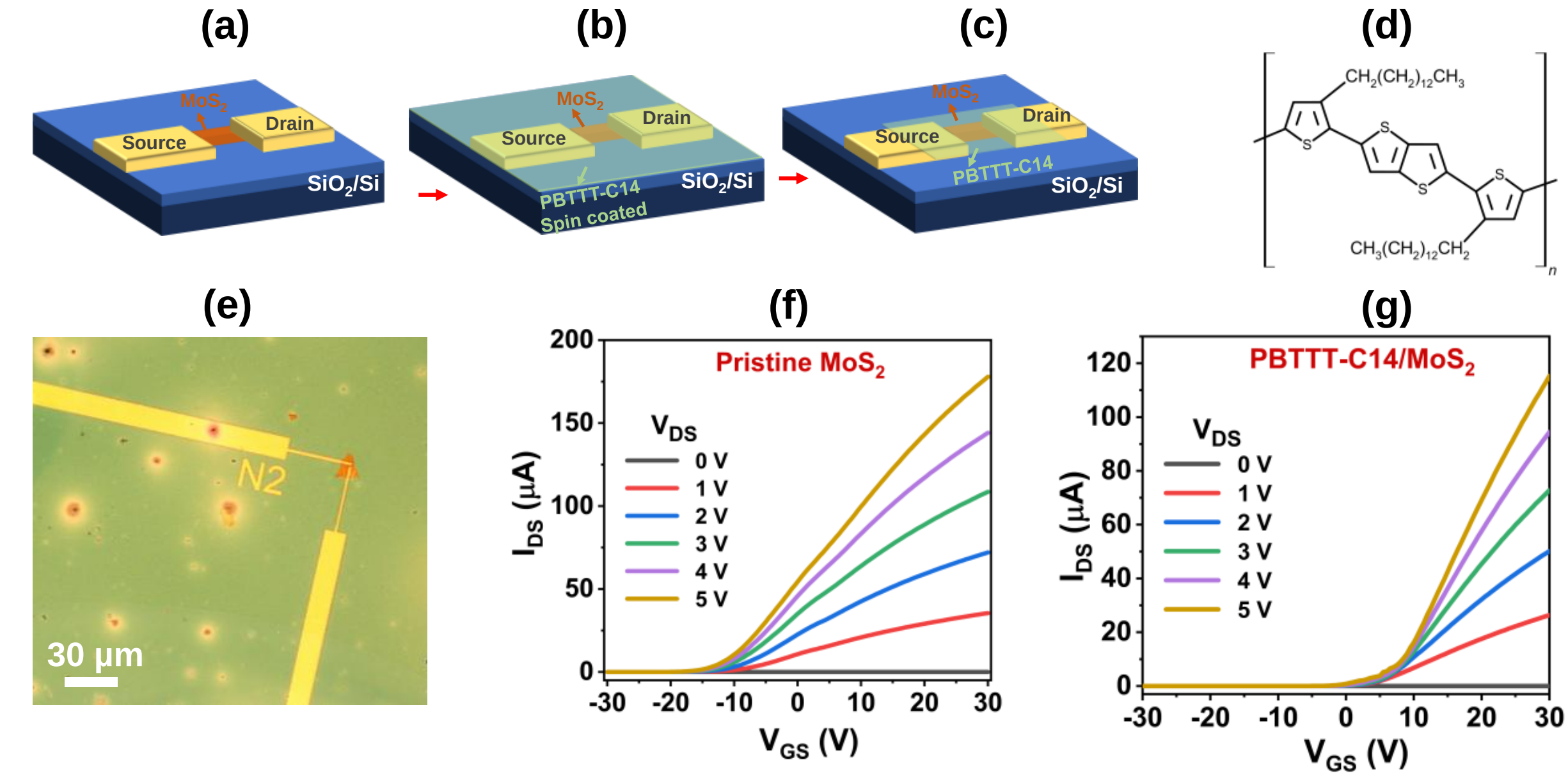

**Figure 1**. (a) Schematic of the pristine back-gated multilayer $MoS_2$ FET on Si/$SiO_2$ substrate, (b) PBTTT-C14 polymer spin-coated on the device, (c) removal of PBTTT-C14 from the contact regions using chlorobenzene for electrical measurements. (d) Chemical structure of PBTTT-C14, (e) Optical image of the fabricated device, (f) Transfer characteristics of the pristine $MoS_2$ FET, showing n-type conduction with depletion-mode (D-mode) operation ($V_{th}$ ≈ –9.6 V, at $V_{DS}$ = 2 V). (g) Transfer characteristics of the PBTTT-C14 coated $MoS_2$ FET, retaining n-type conduction with a strong positive $V_{th}$ shift ($V_{th}$ ≈ +5.9 V, at $V_{DS}$ = 2 V), enabling enhancement-mode operation (E-mode).

The transfer characteristics of the multilayer $MoS_2$ FET before and after PBTTT-C14 coating are shown in Figure 1(f) and 1(g), respectively. The device exhibits typical n-type conduction in both cases, indicating that the PBTTT-C14 overlayer does not alter the majority carrier type of $MoS_2$, and the channel remains electron-transport dominated. The most prominent modification readily observed is a clear positive shift in the threshold voltage ($V_{th}$). To quantify this shift, $V_{th}$ was extracted using the well-established extrapolation in the linear region (ELR) method[46], where the linear extrapolation is taken from the point of maximum transconductance ($g_m$). For the pristine device, $V_{th}$ is −9.6 V (extracted at a drain bias, $V_{DS}$ = 2 V), indicating depletion-mode (D-mode) operation. After PBTTT-C14 coating, $V_{th}$ shifts to +5.9 V, converting the device to enhancement-mode (E-mode) operation. Here, a total positive $V_{th}$ shift of approximately +15.5 V is achieved, primarily driven by interfacial charge-transfer between the PBTTT-C14 overlayer and the $MoS_2$ channel. The detailed mechanism, including the role of polymer energy-level alignment and its effect on the free electron density in $MoS_2$, will be discussed in the subsequent sections. Next, to verify the reliability and repeatability of this modulation, three key checks were performed. First, to rule out short-term transient effects, a PBTTT-C14-coated device was monitored under ambient conditions for 13 days, with periodic measurements of the transfer characteristics. No significant reversal was observed, as shown in Figure 2(a), confirming that the shift is stable and not a temporary electrostatic effect. Second, during the spin coating of PBTTT-C14, the device was annealed at 150 °C for one hour (as described in the Experimental Section). To verify if the observed $V_{th}$ shift is not partially caused by annealing, a control $MoS_2$ FET was annealed under identical conditions without PBTTT-C14. No significant shift in the transfer characteristics was observed, as shown in Figure 2(b), confirming that the positive $V_{th}$ shift arises solely from the PBTTT-C14 layer, Third, to check reproducibility, the same experiment was performed on ten other devices, all of which exhibited a consistent large $V_{th}$ shift and transition from D-mode to E-mode, as shown in Figure 2(c). Please note that the exact $V_{th}$ values vary between devices depending on the $MoS_2$ layer number.

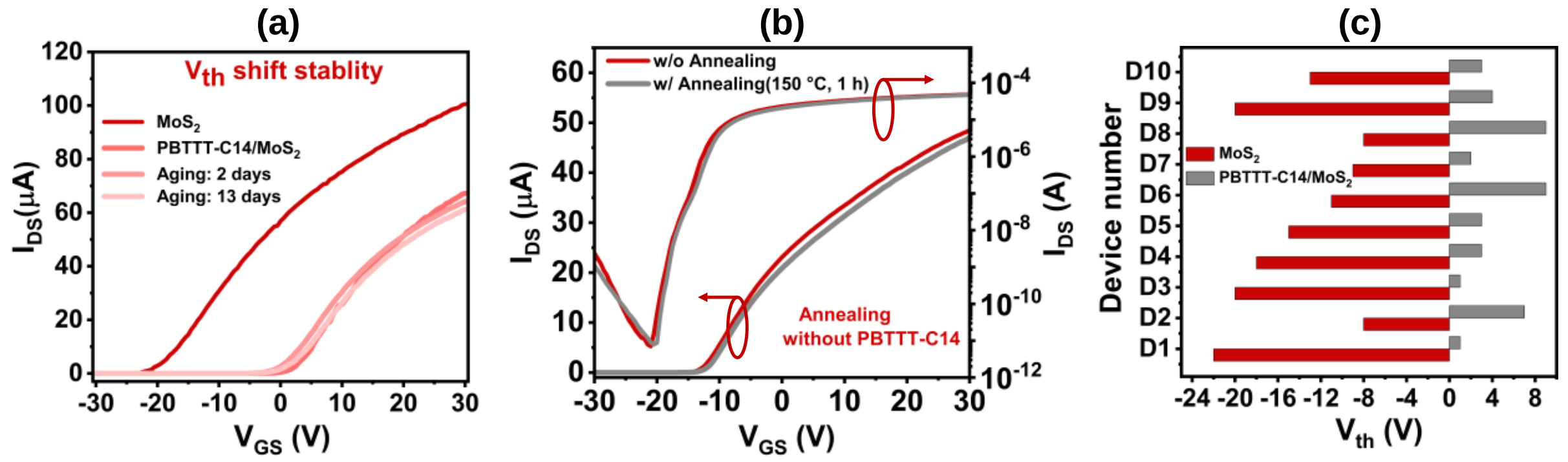

**Figure 2.** (a) Stability of the positive $V_{th}$ shift in PBTTT-C14 coated $MoS_2$ FET monitored over 13 days under ambient conditions (at $V_{DS}$ = 2 V), (b) Transfer characteristics of a control $MoS_2$ FET annealed at 150 °C for 1 h without PBTTT-C14 (at $V_{DS}$ = 2 V), showing no significant $V_{th}$ shift, ensuring that annealing alone does not induce any shift, and (c) Reproducibility of the $V_{th}$ shift across ten devices ($V_{DS}$ = 2 V).

After confirming the repeatability and stability of the $V_{th}$ shift, we next analyse the mobility of the device. The mobility (μ) is determined using the relation: $\mu = g_m \frac{L}{W C_{ox} V_d}$; where L = 3 μm and W = 3 μm are the channel length and width respectively, $C_{ox}$ =1.73×10$^{-8}$ Fcm$^{-2}$ is the gate capacitance per unit area, calculated as $C_{ox} = \frac{\varepsilon_0 \varepsilon_r}{d_{ox}}$, with $\varepsilon_0$ = 8.85×10$^{-12}$ Fm$^{-1}$, $\varepsilon_r$ =3.9 (for $SiO_2$) and $d_{ox}$ = 200 nm. The transconductance $g_m = \frac{dI_{DS}}{dV_{GS}}$ was extracted from the slope of transfer characteristics ($I_{DS}$ vs. $V_{GS}$), and the maximum value of $g_m$ was used to calculate mobility. The mobility remained nearly unchanged after PBTTT-C14 coating, with μ ≈ 54 cm²/V·s for the pristine device and μ ≈ 52 cm²/V·s post-modification, indicating that the PBTTT-C14 layer does not introduce any significant scattering centres.

Furthermore, we observed that the off-current ($I_{off}$) of the treated device is comparatively higher than that of the pristine device. To investigate this, the PBTTT-C14 layer was spin-coated multiple times. As shown in Figure 3(a), increasing the number of coatings gradually decreases $I_{off}$, suggesting that a thin PBTTT-C14 layer may leave microscopic gaps that create unwanted leakage paths. Secondly, this residual $I_{off}$ could be further readily minimized by patterning the PBTTT-C14 layer solely over the channel using a second lithography step, since in the current process the polymer also covers the contacts. However, due to fabrication constraints, these additional optimization steps could not be implemented in the current work.

We next evaluate device hysteresis. Hysteresis is the dependence of the drain current ($I_{DS}$) on the history of the gate voltage ($V_{GS}$), resulting in different $I_{DS}$ vs. $V_{GS}$ paths for forward and backward sweeps[23]. Minimizing hysteresis is critical for reliable device operations. After PBTTT-C14 coating, hysteresis was found to be strongly suppressed, as shown in Figure 3(b). Quantitatively, hysteresis is calculated as the maximum difference in $V_{GS}$ at the same $I_{DS}$ between the forward and backward sweeps in the log-scale transfer curve. The pristine device exhibited a maximum hysteresis of 8.8 V, whereas the PBTTT-C14/$MoS_2$ FET showed a maximum hysteresis of 1.3 V after PBTTT-C14 coating, corresponding to an ~ 85 % reduction. In $MoS_2$ FETs, hysteresis typically arises from charge trapping/detrapping processes associated with surface adsorbates that can bind to defect sites, interfacial traps and defects within the channel[24–26]. Intrinsic sulfur vacancies in $MoS_2$ introduce donor-like defect states[22] that act as charge traps and preferential adsorption sites for ambient species[47], strongly perturbing the channel electrostatics. Such defects have been identified as a key contributor to hysteresis in $MoS_2$ FETs[45]. The pronounced reduction in hysteresis after PBTTT-C14 coating can be explained as follows. First, the polymer forms a van der Waals interface with $MoS_2$[37], creating a clean interface that does not introduce additional trap states or disturb the channel electrostatics. Such an interface enables efficient interfacial charge transfer when the Fermi-level alignment is favourable. Owing to the favourable energy-level alignment[41], between p-type PBTTT-C14 and n-type $MoS_2$, significant charge transfer occurs from $MoS_2$ to the polymer, extracting electrons associated with donor-like sulfur vacancies[48], effectively leading to electrical deactivation of these defects. Park *et al.*[49] demonstrated that charge-transfer doping at a van der Waals organic/$MoS_2$ interface electronically deactivates sulfur-vacancy defect states, which is consistent with our observed hysteresis suppression. This underscores the effectiveness of PBTTT-C14 in enabling both E-mode operation and highly stable device characteristics.

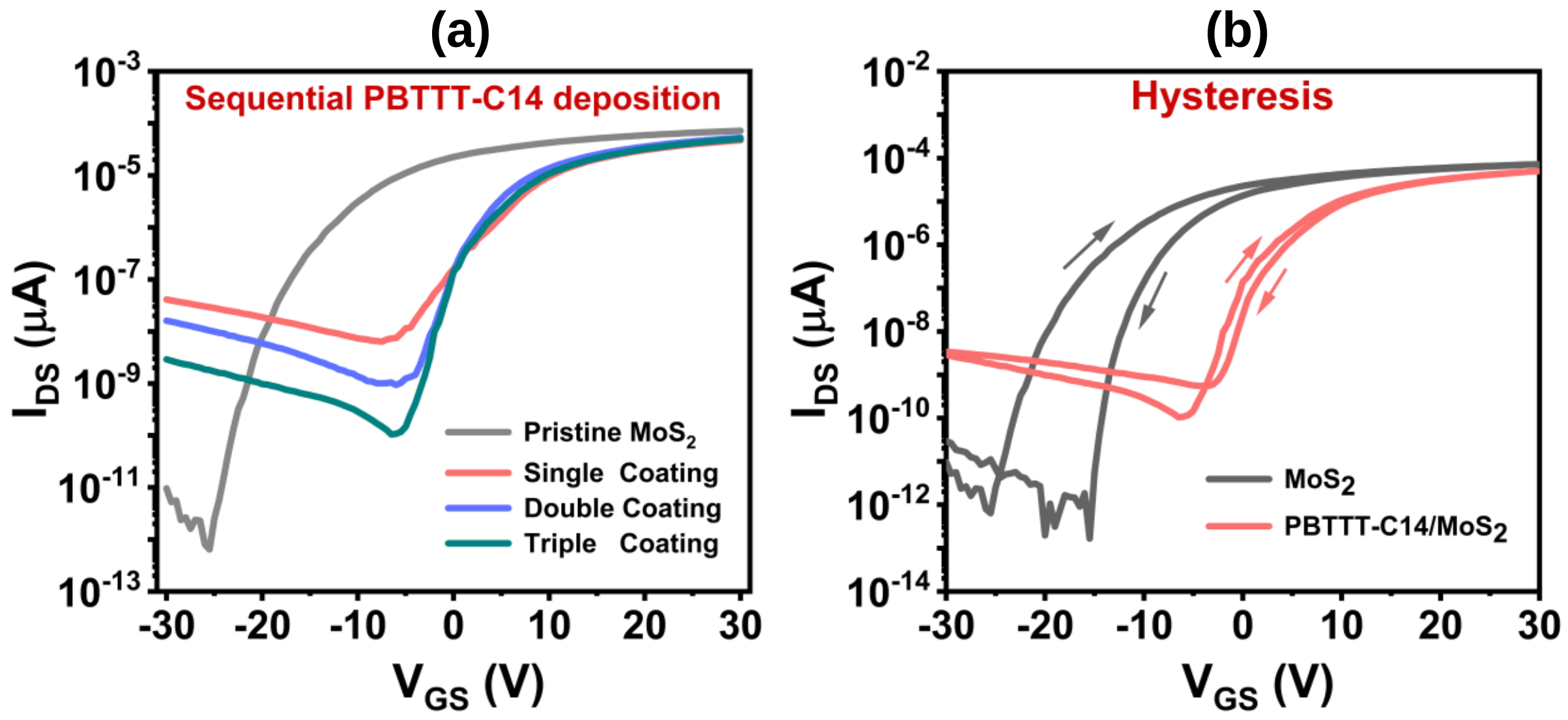

**Figure 3.** (a) Variation of the off-current ($I_{off}$) with PBTTT-C14 layer thickness, showing that increasing the polymer thickness reduces $I_{off}$, (b) Hysteresis in the transfer characteristics before and after PBTTT-C14 coating, demonstrating large suppression of hysteresis (~ 85 %) after surface treatment.

To further elucidate the role of charge transfer and Fermi level alignment, we next investigated two additional organic polymers. The first, P3HT (poly(3-hexylthiophene-2,5-diyl)), is a p-type conjugated polymer with energy levels comparable to PBTTT-C14, while the second, N2200 (Poly{[N,N′-bis(2-octyldodecyl)-naphthalene-1,4,5,8-bis(dicarboximide)-2,6-diyl]-alt-5,5′-(2,2′-bithiophene)}), is an n-type conjugated polymer with a LUMO (lowest unoccupied molecular orbital) level (~ −4.1 eV)[50] closely aligned with the conduction band of n-type $MoS_2$ (~ −4.0 eV)[51], resulting in a small energy offset and minimal charge transfer. Following the same procedure used for PBTTT-C14, P3HT was spin-coated onto the $MoS_2$ FET. The chemical structure of P3HT and the optical image of the device are shown in Figures 4(a) and 4(b), respectively. Similar to PBTTT-C14, it induced a pronounced positive $V_{th}$ shift from −2.1 to +14.5 V (total shift ≈16.6 V), converting the device to E-mode operation as shown in Figure 4(c). Hysteresis was almost completely suppressed, as shown in Figure 4(d). Together, PBTTT-C14 and P3HT have deeper Fermi levels that readily withdraw electrons from $MoS_2$, and induce strong p-type charge transfer doping and strong hysteresis suppression.

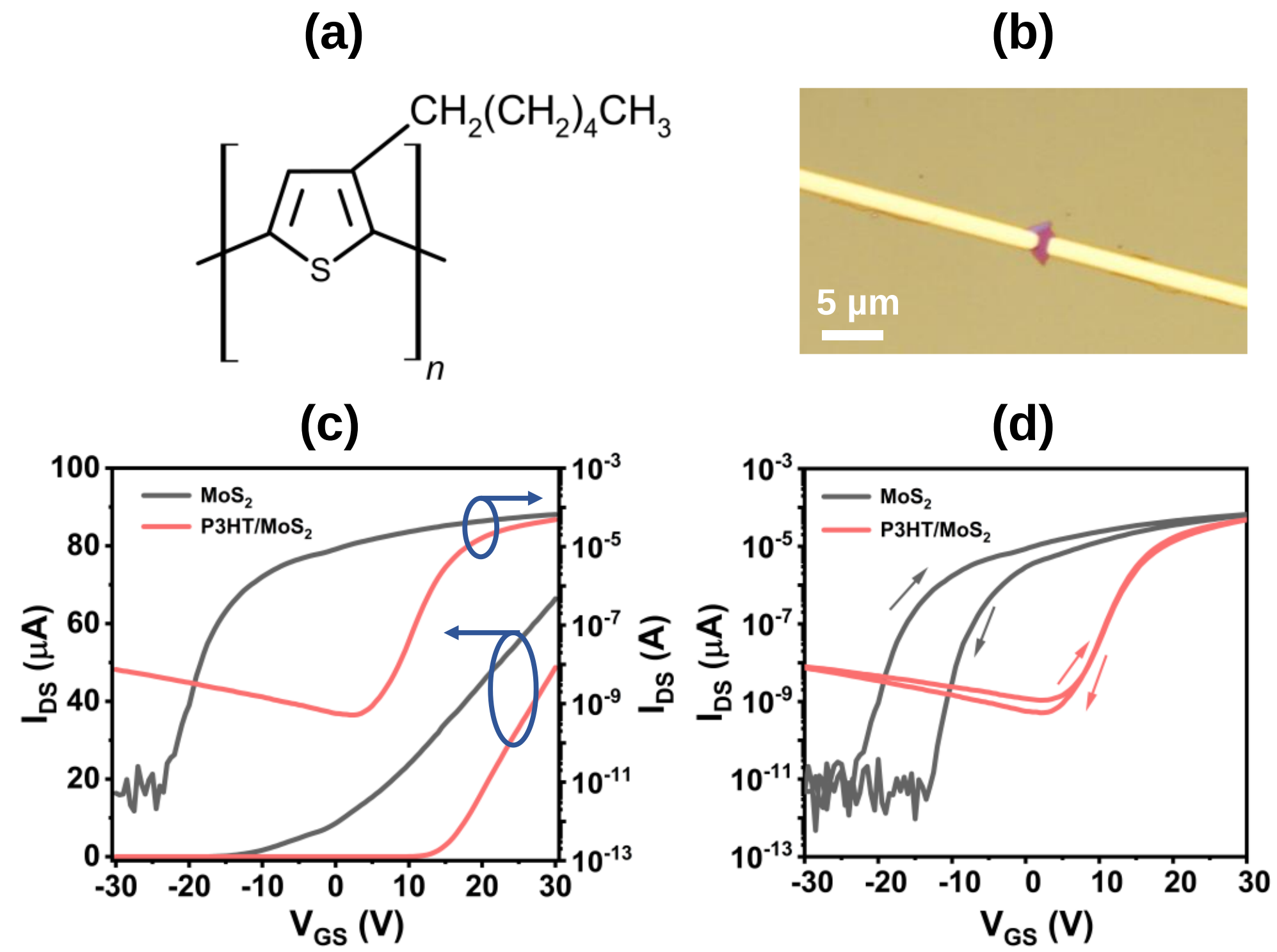

**Figure 4:** (a) Chemical structure of P3HT, (b) Optical image of the fabricated device, (c) Transfer characteristics of the pristine and P3HT-coated $MoS_2$ FETs in linear and log scales, showing a $V_{th}$ shift from −2.1 to +14.5 V at $V_{DS}$ = 5 V and converting from D- mode to E-mode operation, and (d) Double-sweep transfer characteristics of the pristine and P3HT/$MoS_2$ FET, demonstrating hysteresis suppression ($V_{DS}$ = 5 V).

In contrast, coating with N2200 (Chemical structure and optical image of the device is shown in Figures 5(a) and 5(b) respectively), shifted the $V_{th}$ from −13.8 to −8.3 V, corresponding to a total positive shift of only ~5.5 V, and the transistor continued to operate in D-modes as shown in Figure 5(c). Hysteresis was moderately reduced from 11.2 to 5.6 V (~ 50 %), as shown in Figure 5(d). Since both N2200 and $MoS_2$ are n-type and their LUMO and conduction band levels are closely aligned, the energetic offset is insufficient to drive substantial interfacial charge transfer. Please note that the off-current remains low in the N2200-coated device because, being n-type, it does not facilitate conduction at negative gate voltages, leaving any potential leakage paths effectively inactive. Collectively, the behaviours of PBTTT-C14 and P3HT versus N2200 on $MoS_2$ FETs clearly highlight the importance of Fermi-level alignment in governing charge transfer.

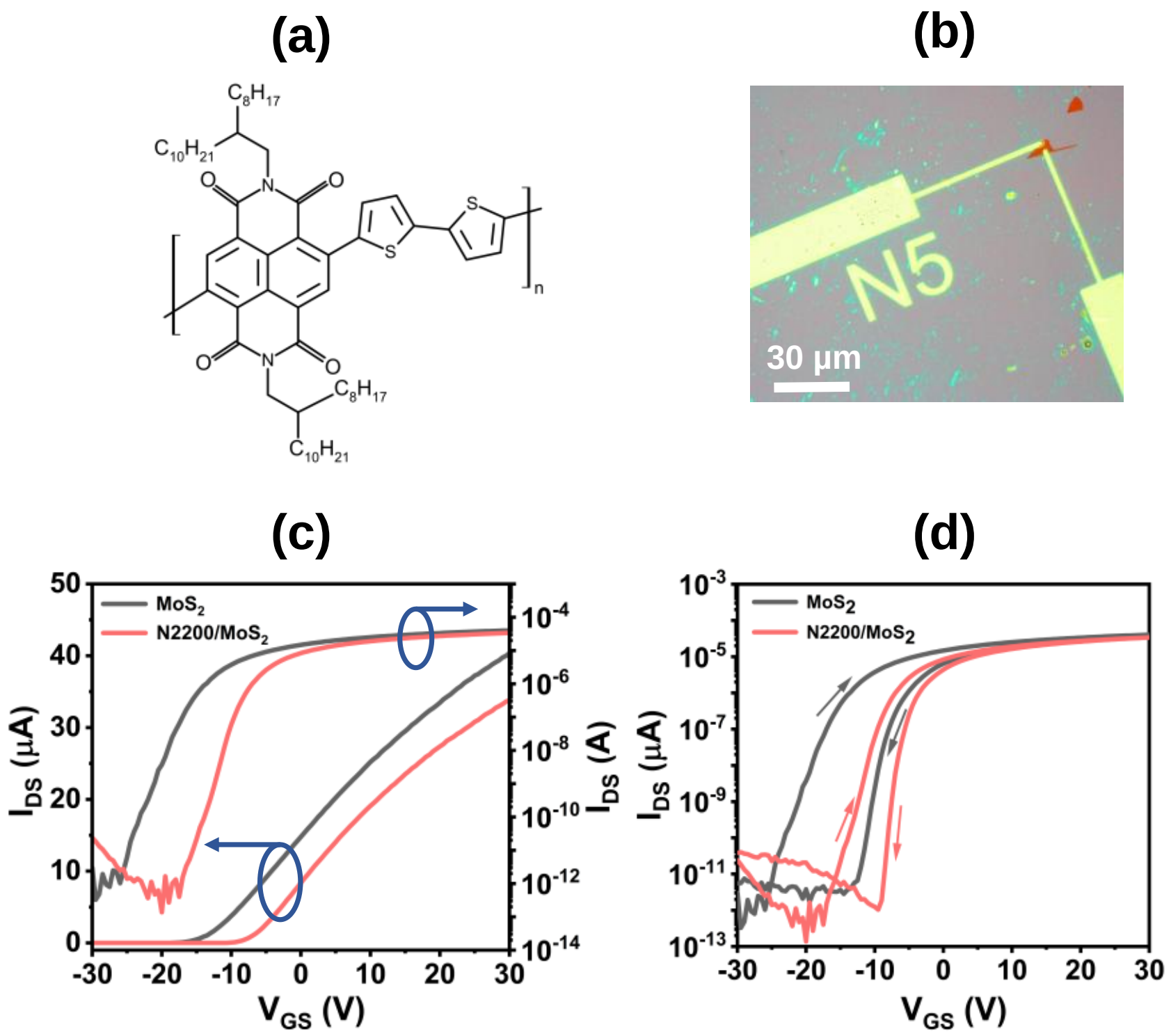

**Figure 5.** (a) Chemical structure of N2200, (b) Optical image of the fabricated device, (c) Transfer characteristics of the pristine and N2200-coated $MoS_2$ FETs in linear and log scales, showing a threshold voltage shift from −13.8 to −8.3 V at $V_{DS}$ = 2 V without achieving E-mode operation (d) Double-sweep transfer characteristics of the pristine and $N2200/MoS_2$ FET, demonstrating hysteresis suppression by ~ 50% ($V_{DS}$ = 2 V).

We next discuss the underlying physics behind this $V_{th}$ modulation. The origin of the negative threshold voltage ($V_{th}$) in pristine $MoS_2$ FETs originates from intrinsic sulfur vacancies, which act as donor-like defects[22] and contribute a high density of free electrons in the channel, rendering the device normally-on[52,53]. Upon coating with the polymer, we expect significant interfacial interactions that reduce the free electron density in the $MoS_2$ channel, effectively lowering the Fermi level. As a result, a higher gate voltage is now required to achieve the onset of conduction, shifting the device toward E-mode operation. Quantitatively, the reduction in free electron density (n) can be found using the relation $\Delta n = C_{\text{ox}}\Delta V_{\text{th}}/q$, where $C_{\text{ox}}$is the gate capacitance per unit area, $q$ is the elementary charge, and $\Delta V_{\text{th}}$is the observed threshold voltage shift. For the PBTTT-C14-coated device, the total reduction in free electron density is approximately $1.67\times10^{12}$ $cm^{-2}$, for P3HT $1.79\times10^{12}$ $cm^{-2}$ while coating with N2200 yields a smaller reduction of $5.93\times10^{11}$ $cm^{-2}$, indicating less electron extraction. This strong modulation of free electron density by PBTTT-C14 and P3HT enables the transition from D-mode to E-mode operation, while N2200 reduces the carrier density only partially and therefore does not induce E-mode operation. Figure 6(a → d) schematically illustrates the relative decrease in free electron density in the $MoS_2$ channel for both polymers. The extraction of free electrons from the $MoS_2$ channel is governed primarily by Fermi-level alignment with the polymer coatings. In PBTTT-C14 and P3HT, the polymer Fermi level lies well below that of $MoS_2$, creating a strong driving force for electron transfer, whereas the near-aligned Fermi level of N2200 results in much weaker extraction. The corresponding band alignment is schematically illustrated in Figure 6(e → h). Specifically, pristine multilayer $MoS_2$ FETs have a conduction band (CB) located at approximately −4.0 eV and a valence band (VB) near −5.2 eV, with the Fermi level positioned close to the CB due to intrinsic n-type doping[51]. PBTTT-C14, a p-type polymer, has HOMO (Highest Occupied Molecular Orbital/LUMO (Lowest Occupied Molecular Orbital) levels at −5.1 eV/−3.1 eV[54], while P3HT exhibits HOMO/LUMO levels at −5.2 eV/−3.2 eV[55]. Since the Fermi levels of these p-type polymers lie close to their HOMO levels, which are substantially deeper than the $MoS_2$ Fermi level, there is a strong energetic driving force for electron transfer from $MoS_2$ into the polymer. This efficient extraction of free electrons lowers the free electron density in the $MoS_2$ channel and shifts the $MoS_2$ Fermi level downward. In contrast, N2200 is an n-type polymer with HOMO/LUMO levels at −5.5 eV/−4.1 eV, and its Fermi level lies close to the LUMO[50].

Because the N2200 LUMO sits only slightly below the $MoS_2$ conduction band edge, the available band offset for electron extraction is small, resulting in much weaker charge transfer. Consequently, the $V_{th}$ shift is modest, and full enhancement-mode operation is not achieved. This highlights how interfacial Fermi-level alignment governs charge transfer, modulates carrier density, and directly influences the threshold voltage. Interfacial charge transfer also plays a critical role in suppressing hysteresis. As discussed earlier, $MoS_2$ FETs exhibit both extrinsic[24] (adsorbates, moisture) and intrinsic (sulfur-vacancy states)[56,57] sources of hysteresis. Extrinsic adsorbates are often bound to sulfur-vacancy sites, meaning that a significant portion of the hysteresis contribution can be traced back to sulfur-vacancy induced trap states[52]. When PBTTT-C14 or P3HT withdraw electrons from the $MoS_2$ channel, these sulfur-vacancy-related traps become electronically passivated. With fewer active trap states available for charge trapping/detrapping, the hysteresis window decreases substantially. This suppression is most pronounced for PBTTT-C14 and P3HT, consistent with their stronger charge-transfer capability compared to N2200.

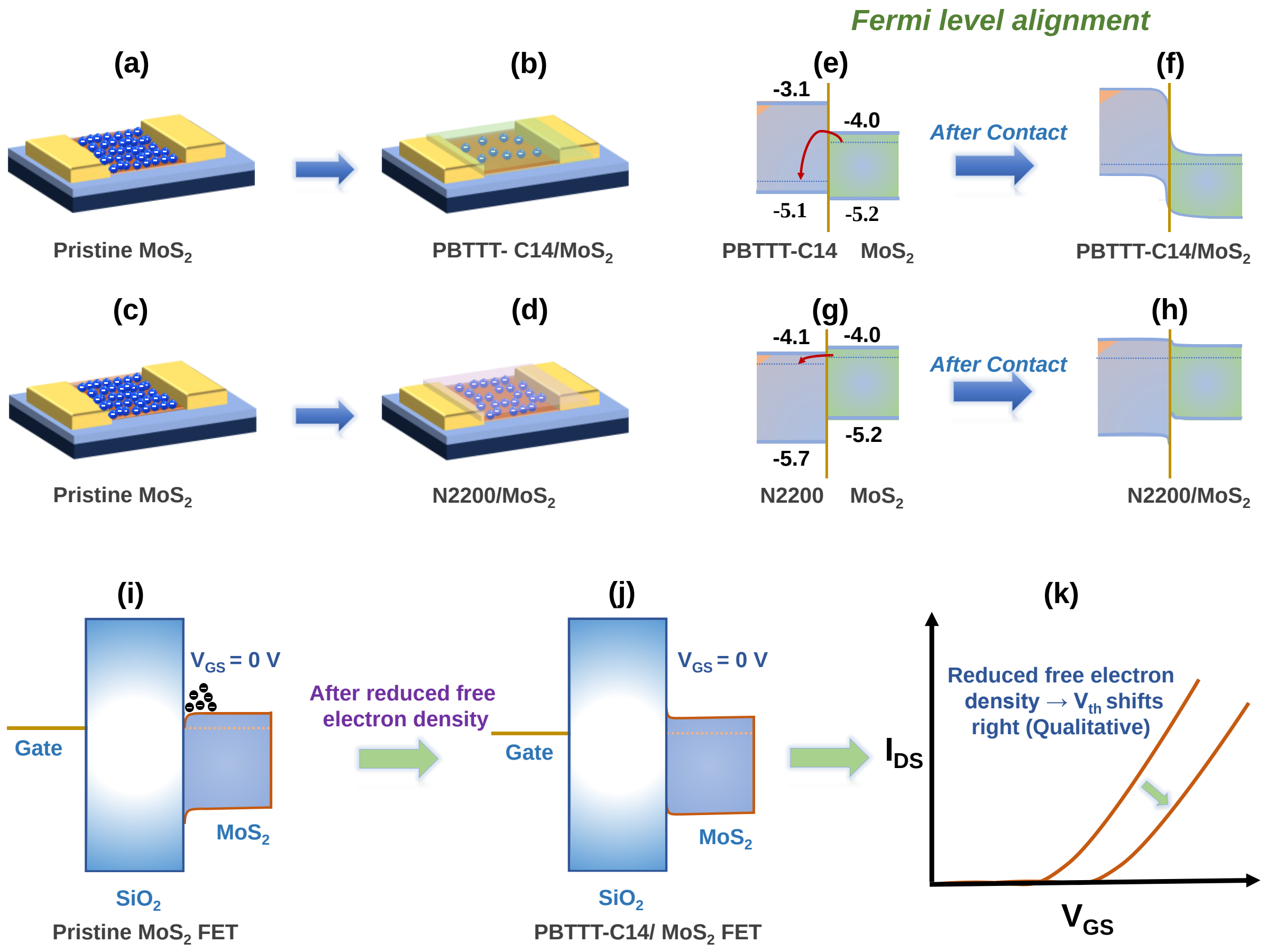

**Figure 6.** **(a → d)** Schematic illustration of the relative reduction of free electrons in the $MoS_2$ channel induced by PBTTT-C14 and N2200 polymer coatings, highlighting the stronger electron extraction achieved with PBTTT-C14. **(e → h)** Schematic illustration of the Band alignment of $MoS_2$ with PBTTT-C14 and N2200. **(i)** Schematic energy-level diagram of a pristine back-gated $MoS_2$ FET at $V_{GS} = 0$ V, where the Fermi level lies close to the conduction band, resulting in finite conduction and depletion-mode operation. **(j)** Effect of PBTTT-C14 coating at $V_{GS} = 0$ V, showing Fermi-level lowering and suppressed channel conduction, thereby requiring a positive gate bias to turn the device on. **(k)** Corresponding electrical manifestation, observed as a strong positive $V_{th}$ shift in $I_{DS}$ Vs. $V_{GS}$ characteristics.

The effect of polymer coating on the energy band diagram at the $MoS_2$/gate interface is schematically illustrated in Figure 6 (i to k). In the pristine $MoS_2$ channel, intrinsic sulfur vacancies contribute abundant free electrons, placing the Fermi level near the conduction band and allowing conduction even at zero gate bias, as depicted in the energy-level diagram in Figure 6(i). Upon PBTTT-C14 coating, interfacial charge transfer extracts free electrons from $MoS_2$, lowering the Fermi level and effectively raising the conduction-band edge, as shown in Figure 6(j). This creates a potential barrier in the channel that must be overcome by a positive gate voltage to enable electron injection. The resulting $I_{DS}$ Vs. $V_{GS}$ characteristics directly reflect this effect through a pronounced positive shift of the threshold voltage, schematically shown in Figure 6(k). To further confirm the role of electronic coupling, an insulating polymer, polydimethylsiloxane (PDMS), was deposited onto the $MoS_2$ FET. The chemical structure of PDMS is shown in Figure 7(a), and the optical image of the device is shown in Figure 7(b). After PDMS deposition, the device exhibits no significant shift in $V_{th}$ (Figure 7(c)) and no appreciable change in hysteresis (Figure 7(d)). These results confirm that mere physical encapsulation does not lead to significant modulation of the electrical properties, underscoring the necessity of interfacial electronic interaction for effective device modulation. Table 1 summarizes the $V_{th}$ modulation and hysteresis behaviour of $MoS_2$ FETs under the various polymer coatings discussed above. Table 2 summarizes reported strategies for tuning threshold voltage and controlling hysteresis in $MoS_2$ FETs. In summary, the study demonstrates that controlled interfacial charge transfer from $MoS_2$ to p-type polymers (PBTTT-C14, P3HT), driven by Fermi level alignment, enables large and stable positive $V_{th}$ shifts, conversion to enhancement-mode operation, and significant hysteresis suppression. These results underscore the potential of organic-inorganic heterostructures for tunable, high-performance 2D transistor technologies, leveraging a dual mechanism of physical passivation and interfacial electronic coupling.

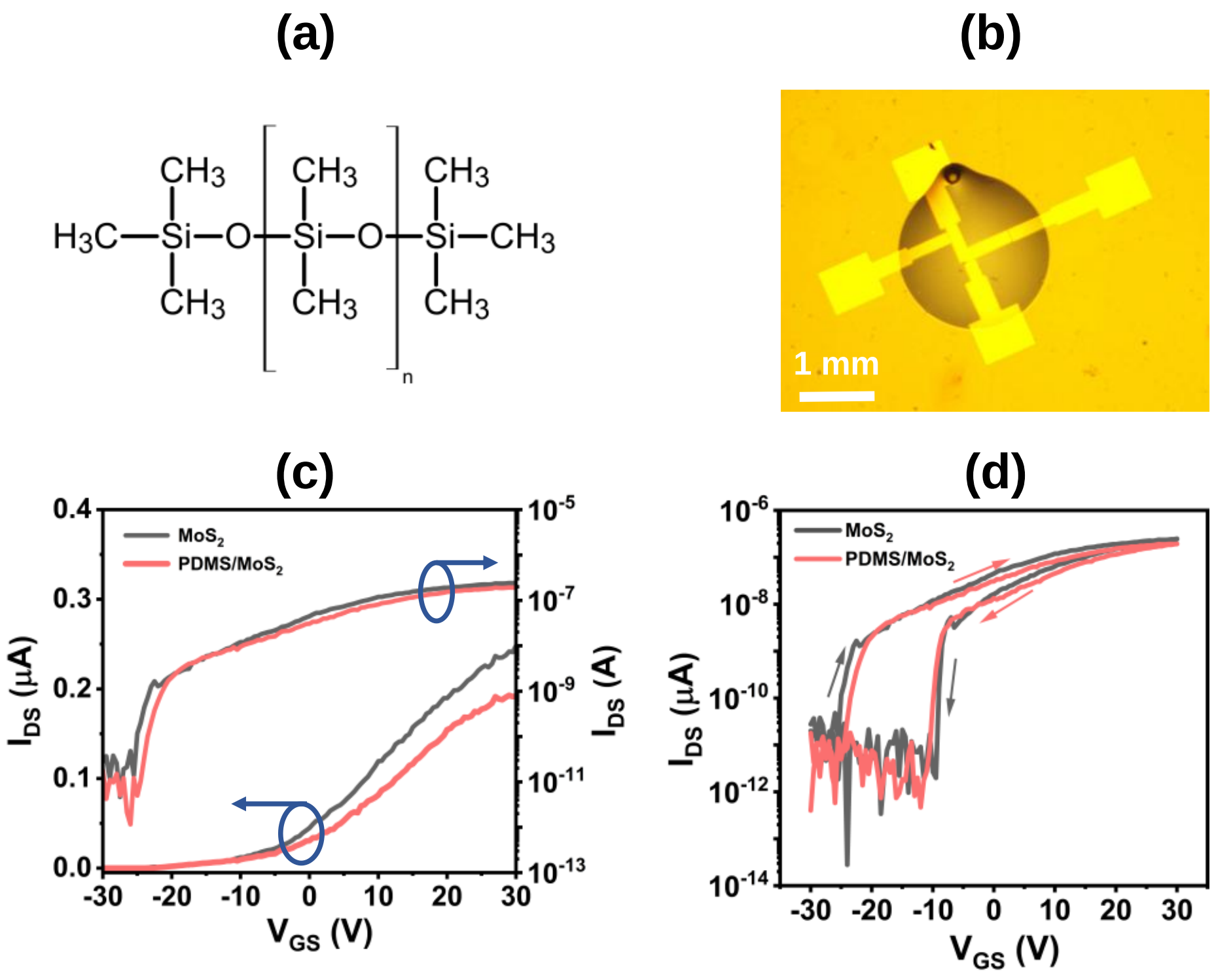

**Figure 7.** (a) Chemical structure of PDMS (Polydimethylsiloxane), (b) Optical image of the device, (c) Transfer characteristics of the pristine and PDMS-coated $MoS_2$ FETs in linear and log scales, showing no significant change in threshold voltage ($V_{th}$) at $V_{DS}$ = 1 V, (d) Double-sweep transfer characteristics of the pristine and PDMS-coated $MoS_2$ FETs (at $V_{DS}$ = 1 V).

**TABLE 1.** Summary of polymer coating -induced electrical modifications in $MoS_2$ FETs.

| Polymer | Polymer-Type | $V_{th}$ (Pristine, V) | $V_{th}$ (Coated, V) | $\Delta V_{th}$ (V) | Operation mode | Hysteresis (Pristine → Coated, V) |
|---|---|---|---|---|---|---|
| PBTTT-C14 | p-type | -9.6 | +5.9 | +15.5 | D → E | 8.8 → 1.3 (~85%↓) |
| P3HT | p-type | -2.1 | +14.5 | +16.6 | D → E | 9.5 → 0 (~100%↓) |
| N2200 | n-type | -13.8 | -8.3 | +5.5 | D → D | 11.2 → 5.6 (~50%↓) |
| PDMS | insulating | -5 | -5 | 0 | D → D | 16.1 → 13.6 (~15%↓) |

**TABLE 2.** Comparison of reported strategies for tuning threshold voltage and hysteresis control in $MoS_2$ FETs.

| Method | Material / Modifier Used | Shift in $V_{th}$ /$V_{turn}$ | Operation mode | Effect on Hysteresis | Reference |
|---|---|---|---|---|---|
| Surface modification with SAM | Octadecyltrichlorosilane (OTS) | ~ +1 V | D→ E-mode | Pristine device nearly hysteresis-free; OTS shows negligible effect | 31 |
| Polymer dielectric engineering | CYTOP | +5.7 V | D→ E-mode | Not reported | 28 |
| Gate metal work function engineering | Pd as gate metal | ~ +1 V | D→ E-mode | Not reported | 27 |
| Molecular surface decoration | F4TCNQ | Maximum ~ +74 (dynamic shift with $V_d$) | D→ D-mode | Hysteresis increased by ~20 V | 58 |
| Plasma treatment | $NH_3$ plasma | + 1.5 V | E→ E-mode | Not reported | 30 |
| Surface charge-transfer doping | Organic p-dopant layer of Magic blue | ~ +100 V | 0 V → E-mode | Not reported | 59 |
| Plasma treatment | Ar plasma + $O_2$ bath | ~ +40 V | D → E-mode | Not reported | 18 |
| Molecular charge transfer doping | F4TCNQ | > 30 V | D → D-mode | Not reported | 60 |
| Dielectric buffer layer engineering | CYTOP | +9.65 | D→ D-mode | Not reported | 61 |
| Encapsulation | h-BN | ~ +10 V | D → D-mode | Fully hysteresis-free | 19 |
| Interface Passivation | APTES | ~ –2 V | D→ D-mode | 23 → 10.8 (53%) | 33 |
| Spin-coating–assisted precipitation | $CsPbBr_3$ nanoclusters | –2.82 V | E→ E-mode | 29 → 21 (27 %) | 35 |
| Plasma treatment and passivation | $O_2$ plasma (30 s) + $Al_2O_3$ passivation | ~ +5 V | D→ D-mode | 14.6 → 0.8 (94%) | 62 |
| Surface treatment with p- type organic polymers | PBTTT-C14 | +15.5 V | D → E-mode | 8.8 → 1.3 (85%) | This work |

**Conclusion:** In conclusion, we demonstrate that interfacial engineering of multilayer $MoS_2$ FETs with p-type organic polymers PBTTT-C14 and P3HT enables simultaneous (1) modulation of threshold voltage ($V_{th}$) and (2) strong hysteresis suppression. Coating with PBTTT-C14 shifted $V_{th}$ from –9.6 V to +5.9 V, effectively converting normally-on, depletion-mode (D-mode) devices into enhancement-mode (E-mode) operation and reducing hysteresis from 8.8 V to 1.3 V (~85%). This behaviour originates from interfacial charge transfer at the $MoS_2$/polymer interface, driven by favourable energy-level alignment. Consistently, P3HT with comparable energy level to PBTTT-C14, produced similar modulation, whereas the n-type polymer N2200 induced modest $V_{th}$ shifts from –13.8 V to +8.3 V, without converting to E-mode operation and hysteresis from 11.2 V to 5.6 V (~50%), underscoring the critical role of polymer-$MoS_2$ energy-level alignment. Control experiments using insulating PDMS showed no significant change in $V_{th}$ and hysteresis, further confirming that mere physical encapsulation is insufficient and electronic interaction at the interface is essential. The reproducibility and stability of these effects were validated across multiple devices. Overall, our findings establish that tailored organic–inorganic heterostructures provide a robust strategy for $V_{th}$ modulation, hysteresis suppression, and the design of high-performance, low-power 2D electronic devices.

**Acknowledgements:** The authors gratefully acknowledge financial support from the Indian Institute of Technology Bombay (IIT Bombay) and the University Grants Commission (UGC), Government of India, for providing fellowships. The authors also sincerely thank the Centre of Excellence in Nanoelectronics (CEN) and the Industrial Research and Consultancy Centre (IRCC), at IIT Bombay for providing access to device fabrication and characterization facilities.

**Conflict of interest:**

The authors declare no conflict of interest.

**References:**

(1) Waldrop, M. M. The Chips Are down for Moore's Law. *Nature* 2016, *530* (7589), 144–147. https://doi.org/10.1038/530144a.

(2) Young, K. K. Short-Channel Effect in Fully Depleted SOI MOSFETs. *IEEE Trans. Electron Devices* 1989, *36* (2), 399–402. https://doi.org/10.1109/16.19942.

(3) Liu, C.; Chen, H.; Wang, S.; Liu, Q.; Jiang, Y.-G.; Zhang, D. W.; Liu, M.; Zhou, P. Two-Dimensional Materials for next-Generation Computing Technologies. *Nat. Nanotechnol.* 2020, *15* (7), 545–557. https://doi.org/10.1038/s41565-020-0724-3.

(4) Schwierz, F. Graphene Transistors. *Nat. Nanotechnol.* 2010, *5* (7), 487–496. https://doi.org/10.1038/nnano.2010.89.

(5) Novoselov, K. S.; Geim, A. K.; Morozov, S. V.; Jiang, D.; Zhang, Y.; Dubonos, S. V.; Grigorieva, I. V.; Firsov, A. A. Electric Field Effect in Atomically Thin Carbon Films. *Science (1979).* 2004, *306* (5696), 666–669. https://doi.org/10.1126/science.1102896.

(6) Kim, K.; Choi, J.-Y.; Kim, T.; Cho, S.-H.; Chung, H.-J. A Role for Graphene in Silicon-Based Semiconductor Devices. *Nature* 2011, *479* (7373), 338–344. https://doi.org/10.1038/nature10680.

(7) Jariwala, D.; Sangwan, V. K.; Lauhon, L. J.; Marks, T. J.; Hersam, M. C. Emerging Device Applications for Semiconducting Two-Dimensional Transition Metal Dichalcogenides. *ACS Nano* 2014, *8* (2), 1102–1120. https://doi.org/10.1021/nn500064s.

(8) Wang, Q. H.; Kalantar-Zadeh, K.; Kis, A.; Coleman, J. N.; Strano, M. S. Electronics and Optoelectronics of Two-Dimensional Transition Metal Dichalcogenides. *Nat. Nanotechnol.* 2012, *7* (11), 699–712. https://doi.org/10.1038/nnano.2012.193.

(9) Patra, U.; Saha, S.; Das, S.; Raturi, M.; Biswas, A.; Sahu, B. P.; Lodha, S.; Dhar, S. Understanding the Role of Environment and Dielectric Capping on the Electrical Properties of $WS_2$ Monolayers Grown by Chemical Vapor Deposition Technique. *J. Appl. Phys.* 2025, *138* (17). https://doi.org/10.1063/5.0294595.

(10) Venkata Subbaiah, Y. P.; Saji, K. J.; Tiwari, A. Atomically Thin $MoS_2$ : A Versatile Nongraphene 2D Material. *Adv. Funct. Mater.* 2016, *26* (13), 2046–2069. https://doi.org/10.1002/adfm.201504202.

(11) Bhatia, S.; Bisht, R. S.; Ahmed, R.; Kumar, P. Trade-Off in Key Electrical Parameters of $MoS_2$ Field-Effect Transistors with the Number of Layers. *ACS Appl. Electron. Mater.* 2025, *7* (15), 6891–6897. https://doi.org/10.1021/acsaelm.5c00734.

(12) Ghosh, S.; Laishram, A. D.; Kumar, P. Simulation on the Miniaturization and Performance Improvement Study of Gr/$MoS_2$ Based Vertical Field Effect Transistor. *Adv. Theory Simul.* 2025, *8* (9). https://doi.org/10.1002/adts.202500139.

(13) Pollmann, E.; Sleziona, S.; Foller, T.; Hagemann, U.; Gorynski, C.; Petri, O.; Madauß, L.; Breuer, L.; Schleberger, M. Large-Area, Two-Dimensional $MoS_2$ Exfoliated on Gold: Direct Experimental Access to the Metal–Semiconductor Interface. *ACS Omega* 2021, *6* (24), 15929–15939. https://doi.org/10.1021/acsomega.1c01570.

(14) Patra, U.; Mujeeb, F.; K, A.; Israni, J.; Dhar, S. Controlled Growth of Millimeter-Size Continuous Bilayer $MoS_2$ Films on $SiO_2$ Substrates by Chemical Vapour Deposition Technique. *Surfaces and Interfaces* 2025, *58*, 105825. https://doi.org/10.1016/j.surfin.2025.105825.

(15) Chuang, H.-J.; Chamlagain, B.; Koehler, M.; Perera, M. M.; Yan, J.; Mandrus, D.; Tománek, D.; Zhou, Z. Low-Resistance 2D/2D Ohmic Contacts: A Universal Approach to High-Performance $WSe_2$ , $MoS_2$ , and $MoSe_2$ Transistors. *Nano Lett.* 2016, *16* (3), 1896–1902. https://doi.org/10.1021/acs.nanolett.5b05066.

(16) Yan, H.; Wang, Y.; Li, Y.; Phuyal, D.; Liu, L.; Guo, H.; Guo, Y.; Lee, T.-L.; Kim, M.; Jeong, H. Y.; Chhowalla, M. A Clean van Der Waals Interface between the High-k Dielectric Zirconium Oxide and Two-Dimensional Molybdenum Disulfide. *Nat. Electron.* 2025, *8* (10), 906–912. https://doi.org/10.1038/s41928-025-01468-1.

(17) Leong, W. S.; Li, Y.; Luo, X.; Nai, C. T.; Quek, S. Y.; Thong, J. T. L. Tuning the Threshold Voltage of $MoS_2$ Field-Effect Transistors via Surface Treatment. *Nanoscale* 2015, *7* (24), 10823–10831. https://doi.org/10.1039/C5NR00253B.

(18) Rai, A. K.; Shah, A. A.; Dar, A. B.; Kumar, J.; Shrivastava, M. Reconfiguration of Intrinsic Depletion-Mode Characteristics of $MoS_2$ Field-Effect Transistors to High-Performance Enhancement-Mode Operation Using an Argon Plasma-Induced P-Type Doping Technique. *Small Methods* 2025, *9* (3). https://doi.org/10.1002/smtd.202401001.

(19) Liu, S.; Yuan, K.; Xu, X.; Yin, R.; Lin, D.; Li, Y.; Watanabe, K.; Taniguchi, T.; Meng, Y.; Dai, L.; Ye, Y. Hysteresis-Free Hexagonal Boron Nitride Encapsulated 2D

Semiconductor Transistors, NMOS and CMOS Inverters. *Adv. Electron. Mater.* 2019, *5* (2). https://doi.org/10.1002/aelm.201800419.

(20) Wani, S.-S.; Hsu, C. C.; Kuo, Y.-Z.; Darshana Kumara Kimbulapitiya, K. M. M.; Chung, C.-C.; Cyu, R.-H.; Chen, C.-T.; Liu, M.-J.; Chaudhary, M.; Chiu, P.-W.; Zhong, Y.-L.; Chueh, Y.-L. Enhanced Electrical Transport Properties of Molybdenum Disulfide Field-Effect Transistors by Using Alkali Metal Fluorides as Dielectric Capping Layers. *ACS Nano* 2024, *18* (16), 10776–10787. https://doi.org/10.1021/acsnano.3c11025.

(21) Younas, R.; Zhou, G.; Hinkle, C. L. A Perspective on the Doping of Transition Metal Dichalcogenides for Ultra-Scaled Transistors: Challenges and Opportunities. *Appl. Phys. Lett.* 2023, *122* (16). https://doi.org/10.1063/5.0133064.

(22) Qiu, H.; Xu, T.; Wang, Z.; Ren, W.; Nan, H.; Ni, Z.; Chen, Q.; Yuan, S.; Miao, F.; Song, F.; Long, G.; Shi, Y.; Sun, L.; Wang, J.; Wang, X. Hopping Transport through Defect-Induced Localized States in Molybdenum Disulphide. *Nat. Commun.* 2013, *4* (1), 2642. https://doi.org/10.1038/ncomms3642.

(23) Late, D. J.; Liu, B.; Matte, H. S. S. R.; Dravid, V. P.; Rao, C. N. R. Hysteresis in Single-Layer $MoS_2$ Field Effect Transistors. *ACS Nano* 2012, *6* (6), 5635–5641. https://doi.org/10.1021/nn301572c.

(24) Shimazu, Y.; Tashiro, M.; Sonobe, S.; Takahashi, M. Environmental Effects on Hysteresis of Transfer Characteristics in Molybdenum Disulfide Field-Effect Transistors. *Sci. Rep.* 2016, *6* (1), 30084. https://doi.org/10.1038/srep30084.

(25) Li, T.; Du, G.; Zhang, B.; Zeng, Z. Scaling Behavior of Hysteresis in Multilayer $MoS_2$ Field Effect Transistors. *Appl. Phys. Lett.* 2014, *105* (9). https://doi.org/10.1063/1.4894865.

(26) Kaushik, N.; Mackenzie, D. M. A.; Thakar, K.; Goyal, N.; Mukherjee, B.; Boggild, P.; Petersen, D. H.; Lodha, S. Reversible Hysteresis Inversion in $MoS_2$ Field Effect Transistors. *NPJ 2D Mater. Appl.* 2017, *1* (1), 34. https://doi.org/10.1038/s41699-017-0038-y.

(27) Wang, H.; Yu, L.; Lee, Y.-H.; Shi, Y.; Hsu, A.; Chin, M. L.; Li, L.-J.; Dubey, M.; Kong, J.; Palacios, T. Integrated Circuits Based on Bilayer $MoS_2$ Transistors. *Nano Lett.* 2012, *12* (9), 4674–4680. https://doi.org/10.1021/nl302015v.

(28) Yoo, G.; Choi, S. L.; Lee, S.; Yoo, B.; Kim, S.; Oh, M. S. Enhancement-Mode Operation of Multilayer $MoS_2$ Transistors with a Fluoropolymer Gate Dielectric Layer. *Appl. Phys. Lett.* 2016, *108* (26). https://doi.org/10.1063/1.4955024.

(29) Leong, W. S.; Li, Y.; Luo, X.; Nai, C. T.; Quek, S. Y.; Thong, J. T. L. Tuning the Threshold Voltage of $MoS_2$ Field-Effect Transistors via Surface Treatment. *Nanoscale* 2015, *7* (24), 10823–10831. https://doi.org/10.1039/C5NR00253B.

(30) Kang, M.; Hong, W.; Lee, I.; Park, S.; Park, C.; Bae, S.; Lim, H.; Choi, S.-Y. Tunable Doping Strategy for Few-Layer $MoS_2$ Field-Effect Transistors via $NH_3$ Plasma

Treatment. *ACS Appl. Mater. Interfaces* 2024, *16* (33), 43849–43859. https://doi.org/10.1021/acsami.4c08549.

(31) Roh, J.; Ryu, J. H.; Baek, G. W.; Jung, H.; Seo, S. G.; An, K.; Jeong, B. G.; Lee, D. C.; Hong, B. H.; Bae, W. K.; Lee, J.; Lee, C.; Jin, S. H. Threshold Voltage Control of Multilayered $MoS_2$ Field-Effect Transistors via Octadecyltrichlorosilane and Their Applications to Active Matrixed Quantum Dot Displays Driven by Enhancement-Mode Logic Gates. *Small* 2019, *15* (7). https://doi.org/10.1002/smll.201803852.

(32) Li, X.; Sun, R.; Guo, H.; Su, B.; Li, D.; Yan, X.; Liu, Z.; Tian, J. Controllable Doping of Transition-Metal Dichalcogenides by Organic Solvents. *Adv. Electron. Mater.* 2020, *6* (3). https://doi.org/10.1002/aelm.201901230.

(33) Han, K. H.; Kim, G.-S.; Park, J.; Kim, S.-G.; Park, J.-H.; Yu, H.-Y. Reduction of Threshold Voltage Hysteresis of $MoS_2$ Transistors with 3-Aminopropyltriethoxysilane Passivation and Its Application for Improved Synaptic Behavior. *ACS Appl. Mater. Interfaces* 2019, *11* (23), 20949–20955. https://doi.org/10.1021/acsami.9b01391.

(34) Liu, N.; Baek, J.; Kim, S. M.; Hong, S.; Hong, Y. K.; Kim, Y. S.; Kim, H.-S.; Kim, S.; Park, J. Improving the Stability of High-Performance Multilayer $MoS_2$ Field-Effect Transistors. *ACS Appl. Mater. Interfaces* 2017, *9* (49), 42943–42950. https://doi.org/10.1021/acsami.7b16670.

(35) Kang, Y. Z.; An, G. H.; Jeon, M.-G.; Shin, S. J.; Kim, S. J.; Choi, M.; Lee, J. B.; Kim, T. Y.; Rahman, I. N.; Seo, H. Y.; Oh, S.; Cho, B.; Choi, J.; Lee, H. S. Increased Mobility and Reduced Hysteresis of $MoS_2$ Field-Effect Transistors via Direct Surface Precipitation of $CsPbBr_3$ -Nanoclusters for Charge Transfer Doping. *Nano Lett.* 2023, *23* (19), 8914–8922. https://doi.org/10.1021/acs.nanolett.3c02293.

(36) Jana, S. P.; Shivangi; Gupta, S.; Gupta, A. K. Enhanced Performance of $MoS_2/SiO_2$ Field-Effect Transistors by Hexamethyldisilazane (HMDS) Encapsulation. *Appl. Phys. Lett.* 2024, *124* (24). https://doi.org/10.1063/5.0204634.

(37) Lee, G.-H.; Lee, C.-H.; van der Zande, A. M.; Han, M.; Cui, X.; Arefe, G.; Nuckolls, C.; Heinz, T. F.; Hone, J.; Kim, P. Heterostructures Based on Inorganic and Organic van Der Waals Systems. *APL Mater.* 2014, *2* (9). https://doi.org/10.1063/1.4894435.

(38) Kang, S. J.; Lee, G.; Yu, Y.; Zhao, Y.; Kim, B.; Watanabe, K.; Taniguchi, T.; Hone, J.; Kim, P.; Nuckolls, C. Organic Field Effect Transistors Based on Graphene and Hexagonal Boron Nitride Heterostructures. *Adv. Funct. Mater.* 2014, *24* (32), 5157–5163. https://doi.org/10.1002/adfm.201400348.

(39) Kumar, P.; Shivananda, K. N.; Zajączkowski, W.; Pisula, W.; Eichen, Y.; Tessler, N. The Relation Between Molecular Packing or Morphology and Chemical Structure or Processing Conditions: The Effect on Electronic Properties. *Adv. Funct. Mater.* 2014, *24* (17), 2530–2536. https://doi.org/10.1002/adfm.201303571.

(40) Bisht, R. S.; Kumar, P. Inverted Resistive Switching Mechanism in Polycrystalline PBTTT-C14 Polymer Devices Based on Contact Geometry and Molecular Packing for

Neuromorphic Memory. *J. Mater. Chem. C Mater.* 2025, *13* (26), 13404–13414. https://doi.org/10.1039/D5TC01372K.

(41) Petoukhoff, C. E.; Kosar, S.; Goto, M.; Bozkurt, I.; Chhowalla, M.; Dani, K. M. Charge Transfer Dynamics in Conjugated Polymer/$MoS_2$ Organic/2D Heterojunctions. *Mol. Syst. Des. Eng.* 2019, *4* (4), 929–938. https://doi.org/10.1039/C9ME00019D.

(42) Jariwala, D.; Howell, S. L.; Chen, K.-S.; Kang, J.; Sangwan, V. K.; Filippone, S. A.; Turrisi, R.; Marks, T. J.; Lauhon, L. J.; Hersam, M. C. Hybrid, Gate-Tunable, van Der Waals p–n Heterojunctions from Pentacene and $MoS_2$. *Nano Lett.* 2016, *16* (1), 497–503. https://doi.org/10.1021/acs.nanolett.5b04141.

(43) Bisht, R. S.; Kumar, P. Simulation Study of Various Factors Affecting the Performance of Vertical Organic Field-Effect Transistors. *Engineering Research Express* 2023, *5* (3), 035040. https://doi.org/10.1088/2631-8695/acf029.

(44) Liu, H.; Liu, D.; Yang, J.; Gao, H.; Wu, Y. Flexible Electronics Based on Organic Semiconductors: From Patterned Assembly to Integrated Applications. *Small* 2023, *19* (11). https://doi.org/10.1002/smll.202206938.

(45) Gu, M.; Kim, T.; Jeon, D.; Lee, D.; Park, J.; Kim, T. Correlation between Fermi-Level Hysteresis and Sulfur Vacancy-Based Traps on $MoS_2$. *ACS Appl. Electron. Mater.* 2024, *6* (11), 8525–8531. https://doi.org/10.1021/acsaelm.4c01808.

(46) Ortiz-Conde, A.; García Sánchez, F. J.; Liou, J. J.; Cerdeira, A.; Estrada, M.; Yue, Y. A Review of Recent MOSFET Threshold Voltage Extraction Methods. *Microelectronics Reliability* 2002, *42* (4–5), 583–596. https://doi.org/10.1016/S0026-2714(02)00027-6.

(47) Al Mamun, M. S.; Sainoo, Y.; Takaoka, T.; Ando, A.; Komeda, T. Hysteresis in the Transfer Characteristics of $MoS_2$ Field Effect Transistors: Gas, Temperature and Photo-Irradiation Effect. *RSC Adv.* 2024, *14* (49), 36517–36526. https://doi.org/10.1039/D4RA04820B.

(48) Li, L.; Long, R.; Bertolini, T.; Prezhdo, O. V. Sulfur Adatom and Vacancy Accelerate Charge Recombination in $MoS_2$ but by Different Mechanisms: Time-Domain Ab Initio Analysis. *Nano Lett.* 2017, *17* (12), 7962–7967. https://doi.org/10.1021/acs.nanolett.7b04374.

(49) Park, J. H.; Sanne, A.; Guo, Y.; Amani, M.; Zhang, K.; Movva, H. C. P.; Robinson, J. A.; Javey, A.; Robertson, J.; Banerjee, S. K.; Kummel, A. C. Defect Passivation of Transition Metal Dichalcogenides via a Charge Transfer van Der Waals Interface. *Sci. Adv.* 2017, *3* (10). https://doi.org/10.1126/sciadv.1701661.

(50) Su, W.; Meng, Y.; Guo, X.; Fan, Q.; Zhang, M.; Jiang, Y.; Xu, Z.; Dai, Y.; Xie, B.; Liu, F.; Zhang, M.; Russell, T. P.; Li, Y. Efficient and Thermally Stable All-Polymer Solar Cells Based on a Fluorinated Wide-Bandgap Polymer Donor with High Crystallinity. *J. Mater. Chem. A Mater.* 2018, *6* (34), 16403–16411. https://doi.org/10.1039/C8TA05376F.

(51) Trainer, D. J.; Putilov, A. V.; Di Giorgio, C.; Saari, T.; Wang, B.; Wolak, M.; Chandrasena, R. U.; Lane, C.; Chang, T.-R.; Jeng, H.-T.; Lin, H.; Kronast, F.; Gray, A. X.; Xi, X.; Nieminen, J.; Bansil, A.; Iavarone, M. Inter-Layer Coupling Induced Valence Band Edge Shift in Mono- to Few-Layer $MoS_2$. *Sci. Rep.* 2017, *7* (1), 40559. https://doi.org/10.1038/srep40559.

(52) Kim, I. S.; Sangwan, V. K.; Jariwala, D.; Wood, J. D.; Park, S.; Chen, K.-S.; Shi, F.; Ruiz-Zepeda, F.; Ponce, A.; Jose-Yacaman, M.; Dravid, V. P.; Marks, T. J.; Hersam, M. C.; Lauhon, L. J. Influence of Stoichiometry on the Optical and Electrical Properties of Chemical Vapor Deposition Derived $MoS_2$. *ACS Nano* 2014, *8* (10), 10551–10558. https://doi.org/10.1021/nn503988x.

(53) Lee, I.; Kang, M.; Park, S.; Park, C.; Lee, H.; Bae, S.; Lim, H.; Kim, S.; Hong, W.; Choi, S. Healing Donor Defect States in CVD-Grown $MoS_2$ Field-Effect Transistors Using Oxygen Plasma with a Channel-Protecting Barrier. *Small* 2024, *20* (2). https://doi.org/10.1002/smll.202305143.

(54) Cochran, J. E.; Junk, M. J. N.; Glaudell, A. M.; Miller, P. L.; Cowart, J. S.; Toney, M. F.; Hawker, C. J.; Chmelka, B. F.; Chabinyc, M. L. Molecular Interactions and Ordering in Electrically Doped Polymers: Blends of PBTTT and F4TCNQ. *Macromolecules* 2014, *47* (19), 6836–6846. https://doi.org/10.1021/ma501547h.

(55) Petoukhoff, C. E.; Kosar, S.; Goto, M.; Bozkurt, I.; Chhowalla, M.; Dani, K. M. Charge Transfer Dynamics in Conjugated Polymer/$MoS_2$ Organic/2D Heterojunctions. *Mol. Syst. Des. Eng.* 2019, *4* (4), 929–938. https://doi.org/10.1039/C9ME00019D.

(56) Cao, B.; Wang, Z.; Xiong, X.; Gao, L.; Li, J.; Dong, M. Hysteresis-Reversible $MoS_2$ Transistor. *New Journal of Chemistry* 2021, *45* (27), 12033–12040. https://doi.org/10.1039/D1NJ01267C.

(57) Shu, J.; Wu, G.; Guo, Y.; Liu, B.; Wei, X.; Chen, Q. The Intrinsic Origin of Hysteresis in $MoS_2$ Field Effect Transistors. *Nanoscale* 2016, *8* (5), 3049–3056. https://doi.org/10.1039/C5NR07336G.

(58) Kii, H.; Nouchi, R. Drain-Induced Threshold-Voltage Shift Greater than 90 V/V in Molecule-Decorated $MoS_2$ Field-Effect Transistors Operated in Air. *ACS Appl. Electron. Mater.* 2025, *7* (11), 5282–5289. https://doi.org/10.1021/acsaelm.5c00627.

(59) Tarasov, A.; Zhang, S.; Tsai, M.; Campbell, P. M.; Graham, S.; Barlow, S.; Marder, S. R.; Vogel, E. M. Controlled Doping of Large-Area Trilayer $MoS_2$ with Molecular Reductants and Oxidants. *Advanced Materials* 2015, *27* (7), 1175–1181. https://doi.org/10.1002/adma.201404578.

(60) Wang, J.; Ji, Z.; Yang, G.; Chuai, X.; Liu, F.; Zhou, Z.; Lu, C.; Wei, W.; Shi, X.; Niu, J.; Wang, L.; Wang, H.; Chen, J.; Lu, N.; Jiang, C.; Li, L.; Liu, M. Charge Transfer within the F 4 TCNQ-$MoS_2$ van Der Waals Interface: Toward Electrical Properties Tuning and Gas Sensing Application. *Adv. Funct. Mater.* 2018, *28* (51). https://doi.org/10.1002/adfm.201806244.

(61) Lee, H. S.; Shin, J. M.; Jeon, P. J.; Lee, J.; Kim, J. S.; Hwang, H. C.; Park, E.; Yoon, W.; Ju, S.; Im, S. Few-Layer $MoS_2$ –Organic Thin-Film Hybrid Complementary Inverter Pixel Fabricated on a Glass Substrate. *Small* 2015, *11* (18), 2132–2138. https://doi.org/10.1002/smll.201402950.

(62) Liu, N.; Baek, J.; Kim, S. M.; Hong, S.; Hong, Y. K.; Kim, Y. S.; Kim, H.-S.; Kim, S.; Park, J. Improving the Stability of High-Performance Multilayer $MoS_2$ Field-Effect Transistors. *ACS Appl. Mater. Interfaces* 2017, *9* (49), 42943–42950. https://doi.org/10.1021/acsami.7b16670.